\documentclass[a4paper,12pt]{article}
\usepackage{graphicx}
\usepackage[dvips]{color}  
\begin{document}

\parindent=0mm
\parskip 1.5mm

\begin{center}
{\LARGE \bf Complex fission phenomena}

\bigskip

\small \it

D. N. Poenaru$^{1,2}$, R. A. Gherghescu$^1$, and W. Greiner$^2$

$^1$Horia Hulubei National Instiute of Physics and Nuclear Engineering,
PO Box MG-6, \\RO-077125 Bucharest-Magurele, Romania\\
$^2$Frankfurt Institute for Advanced Studies, J W Goethe University,
Pf. 111932, \\D-60054 Frankfurt am Main, Germany\\

\end{center}

\bigskip

\normalsize \rm

\begin{abstract}
Complex fission phenomena are studied in a unified way. Very general
reflection asymmetrical equilibrium (saddle point) nuclear shapes are
obtained by solving an integro-differential equation without being necessary
to specify a certain parametrization. The mass asymmetry in binary cold
fission of Th and U isotopes is explained as the result of adding a
phenomenological shell correction to the liquid drop model deformation
energy. Applications to binary, ternary, and quaternary fission are
outlined. 
\end{abstract}

PACS: 24.75.+i, 25.85.Ca, 27.90.+b, 23.70.+j

\bigskip 

{\em Keywords:} NUCLEAR COLD FISSION, multicluster fission, true ternary
fission, binary fission, saddle point shapes, integro-differential equation,
mass asymmetry.

\bigskip \bigskip

{\em E-PRINT at http://arXiv.org:} nucl-th/0404029.

\bigskip \bigskip

\section{Introduction}

There is a continuous progress  in deeper understanding the large variety of
fission processes which can be theoretically treated in a unified way
\cite{p195b96b}. 
Even $\alpha$-decay and cluster radioactivities
\cite{p240pr02,roy01np} can be considered among the members of this family. 
Particle accompanied fission was discovered in 1946 (see the review
\cite{mut96mb}), 
but only recently, by using new methods of fission fragment identification
based on the characteristic rotational spectra measured with large arrays of
Germanium Compton-suppressed detectors (as GAMMASPHERE)
\cite{ham94jpg,ham95ppnp}, the ``cold'' processes $\alpha$ and $^{10}$Be
accompanied cold fission as well as the double and triple fine structure in
a binary and ternary fission have been discovered
\cite{ram98prl}. The unified approach of cold binary fission, cluster
radioactivity, and
$\alpha $-decay \cite{p195b96b} was extended to cold ternary 
\cite{p224jpg00} and to multicluster fission including quaternary 
(two-particle accompanied) fission \cite{p218pr99}. In that paper we
stressed the expected enhanced yield of two alpha accompanied fission
compared to other combinations of two light particles; 
it was indeed experimentally confirmed \cite{gon02pc,gon03hip}.
In a cold binary fission the involved nuclei 
are neither excited nor strongly deformed, hence no neutron is evaporated 
from the fragments or from the compound nucleus; the total kinetic energy 
equals the released energy. In a more
complex than binary {\em cold fission} (ternary, quaternary, etc),  neutrons 
could still be emitted from the neck, because the $Q$-value is positive. In 
this case their kinetic energy added to those of the fragments should exhaust
the total released energy.

The importance of scission configuration for ternary fission
\cite{gav75pr} was repeatedly stressed in the past.
In binary fission, it is now better understood due to a longstanding effort
of systematic analysis \cite{nag96pl,nis97prc,oht99gb,zha99prl}.

The statical approach was widely used \cite{coh63ap,str62jetf,has88b} to
find the saddle point shapes within a liquid drop model (LDM). Usually the
equilibrium nuclear shapes are obtained by minimizing the energy functional
on a certain class of trial functions representing the surface equation.
Such an approach shows the importance of taking into account a large number
of deformation coordinates (it seems that 5 coordinates are frequently
needed) \cite{smo93app,mol01n}. The parametrization of Legendre polynomial
expansion with even order deformation parameters $\alpha _{2n}$ up to $n=18$
was employed \cite{coh63ap} to describe various saddle point shapes
including those very similar to two tangent spheres.
 
In order to study nuclear properties we have to consider both the collective
and the single-particle motion of nucleons. This can be done by adding a
shell correction to the LDM deformation energy \cite{str67np}.
Otherwise a well known asymmetrical mass distribution of fission fragments
or the ground state deformation of the majority of nuclei could not be
explained. 

In this paper we present results obtained with a method
allowing to find a general reflection symmetrical or asymmetrical
saddle point shape as a solution of an integro-differential equation without
a shape parametrization {\em apriori} introduced.  This equation is derived
as a Euler-Lagrange relationship associated to the variational problem of
minimizing the potential energy with constraints (constant volume and given
deformation parameter). The axially-symmetrical surface shape minimizing the
liquid drop energy, $E_{LDM}=E_s + E_C$, is straightforwardly obtained. 
Minima of the saddle point deformation energy appear at finite values of the
mass-asymmetry parameter as soon as the shell corrections, $\delta E$, are
taken into account \cite{p241jnrs02,p249el}. 
A phenomenological shell correction is used. Also we shall discuss the
multicluster fission phenomena.

\section{Equilibrium shapes}\label{shapes}

In contrast to reality, within a liquid drop model all nuclear shapes in the
ground-state are spherical and the fission fragment mass distributions are
symmetrical. 
Permanent nuclear deformations and fission fragment mass asymmetry can be
explained by combining the collective (liquid drop-like) and single particle
properties in the framework of a macroscopic-microscopic method. By using
the two center shell model to describe the single-particle states in binary
fission or the three center shell model in the ternary fission, one can
follow the shell structure all the way from the original nucleus, over the
potential barriers, up to the final stage of separated fragments.
Particularly important points on a potential energy surface are those
corresponding to the ground-state, saddle point and scission point.

\begin{figure}[ht] \hspace*{4cm}
\includegraphics[width=8.5cm]{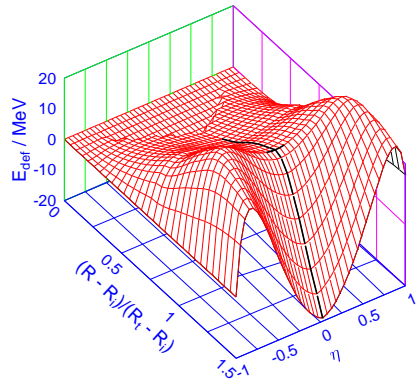} 
 \vspace{-0.8cm}
 \caption{An example of a potential energy surface $E_{def}=E_{def}(R,\eta)$
calculated within liquid drop model,
with a statical path marked with the heavy line which has a cross at the 
saddle point.\label{Fig.6}}
 \end{figure}

In order to illustrate the definition of the saddle point we plotted in
figure~\ref{Fig.6} an example of a LDM potential energy surface
(PES) versus two deformation parameters: the elongation 
$R$ [or the dimensionless quantity  $(R-R_i)/(R_t - R_i)$ where $R_i$ and
$R_t$ are the initial and touching point values of
the separation distance $R$] and the mass asymmetry $\eta
=(A_1-A_2)/(A_1 + A_2)$. A
statical path may be seen on this PES as a heavy line following the valley
of the potential minimum which corresponds to $\eta = 0$. If we start with a
large value of $R$ and then follow the decreasing elongations, the bottom of
that valley leads to increasingly higher energies up to a maximum (the
saddle point) marked with a cross on figure~\ref{Fig.6}, then the energies
along the valley are decreasing until the ground state minimum is reached. 
In this example the energy $E=E(R,\eta)$ is function of two shape
coordinates and the fission valley represents a conditional minimum of the
energy [$\partial E / \partial \eta =0$ for different elongations $R=R_k$ 
($k=1, 2, ..., n$) and $\eta=0$].  The maximum value of this minimum is 
the saddle point defined by 
\begin{equation}
\partial E / \partial \eta = \partial E / \partial R =0
\end{equation}
\begin{equation}
\left| \begin{array}{cc}
\frac{\partial ^2 E}{\partial R ^2} &
\frac{\partial ^2 E}{\partial R \partial \eta} \\
\frac{\partial ^2 E}{\partial \eta   \partial R} &
\frac{\partial ^2 E}{\partial \eta ^2} \end{array} \right| < 0
\end{equation}
If we take another value of mass asymmetry $\eta=\eta_k \neq 0$, for every 
$R_j$ within LDM,  we obtain a new (conditional) saddle point at higher
energy (see figure~\ref{fig13} where $d_L-d_R$ is proportional with
$\eta$), proving that the mass asymmetric distribution of fission fragments
is not explained by a pure LDM.

\section{Integro-differential equation}\label{equ}
 
We are looking for a function $\rho=\rho(z)$ expressing in cylindrical
coordinates the nuclear surface equation with axial symmetry around $z$
axis and the tips $z_1$ and $z_2$. 
The dependence on the neutron, $N$, and proton, $Z$, numbers is
contained in the surface energy of a spherical nucleus, $E_s^0$, the
fissility parameter $X=E_C^0/(2E_s^0)$, as well as in the shell
correction of the spherical nucleus $\delta E^0$.
$E_C^0$ is the Coulomb energy of the spherical shape for which the radius is
$R_0=r_0A^{1/3}$ and the mass number $A=N+Z$. The radius constant is
$r_0=$~1.2249~fm, and $e^2=1.44$~MeV$\cdot$fm is the square of electron
charge. The lengths are given in units of the radius, $R_0$, and the Coulomb
potential at the nuclear surface, $V_s=(R_0/Ze)\phi_s$,  in units of $Ze/R_0$. 
The surface tension and the charge density are denoted by $\sigma$ and
$\rho_e$ respectively.
The nuclear surface equation we are looking for
should minimize the functional of potential energy of deformation
\begin{equation}
E_s + E_C = 2\pi \sigma R_0^2\int_{z_1}^{z_2}\rho(z)\sqrt{1 + \rho'^2}dz
+ \frac {2\pi R_0^2 Ze \rho _e}{5} \int _{z_1} ^{z_2}
\left ( \rho ^2 - \frac {z}{2} \frac {\partial \rho ^2}{\partial z}
\right )V_s dz
\label{eq:a}
\end{equation}
with two constraints: volume conservation, 
\begin{equation}
V=\pi R_0^3\int_{z_1}^{z_2} \rho^2(z)dz = \frac{4\pi R_0^3}{3}
\label{eq:v}
\end{equation}
and a given deformation parameter, 
\begin{equation}
\alpha = \frac{\pi R_0^3}{V}\int_{z_1}^{z_2} F(z,\rho)\rho^2dz
\label{eq:d}
\end{equation}
assumed to be an adiabatic variable.

According to the calculus of variations \cite{arf95b} the function $\rho
(z)$ minimizing the energy with two constraints should satisfy the
Euler-Lagrange equation (see the Appendix) 
\begin{equation}
\rho\rho''- \rho'^2- \left( \lambda_1 + \lambda_2|z| + 
6XV_s \right)\rho (1+\rho'^2)^{3/2} - 1 = 0
\label{eq:sec}
\end{equation}
or
\begin{equation}
2\sigma K + 3\rho_e \phi_s/5 + {\lambda}'_1 + {\lambda}'_2 |z| = 0 
\label{eq:fi}
\end{equation} 
where ${\lambda}'_1$ and ${\lambda}'_2$ are Lagrange multipliers and $K$ is
the mean curvature \cite{web04}:
\begin{equation}
K=({\cal{R}}_1^{-1} + {\cal{R}}_2^{-1})/2
\end{equation}
with ${\cal{R}}_1$ and ${\cal{R}}_2$ the principal radii of curvature given
by
\begin{equation}
{\cal{R}}_1=R_0\tau \rho \; \; ; \; \; {{\cal R}_2} = - R_0\tau^3/\rho''
\; \; ; \; \; \tau^2=1+\rho'^2
\end{equation}
where $\rho'=d\rho/dz$ and $\rho''=d^2\rho/dz^2$. 

It is interesting to mention that in the absence of an electric charge, the
condition of stable equilibrium at the surface of a fluid
\cite{lam32b,lan59b} is given by Laplace formula equating the difference of
pressures with the product $2\sigma K$.

By choosing the deformation coordinate as the distance between the centers
of mass of the left and right fragments, $\alpha = |z^c_L|+|z^c_R|$, one can
reach all intermediate stages of deformation from one parent nucleus to two
fragments by a continuos variation of its value. Also a possible dynamical
study, for which the center of mass treatment is very important
\cite{p150zp89}, may conveniently use this definition of the deformation
parameter. The position of separation plane between fragments, $z=0$, is
given by the condition
$\left(d\rho/dz\right)_{z=0}=0$, which defines the median plane for a usual
spherical, ellipsoidal, or ``diamond'' shape in the ground state, or the
middle of the neck for an elongated reflection symmetrical shape on the
fission path. For this choice of the function $F(z,\rho)$ one has $f=|z|$.

At the left hand side and right hand side tips on the symmetry axis 
one can write 
\begin{equation}
\rho(z_1)=\rho(z_2)=0
\end{equation}
and the transversality conditions
\begin{equation}
\frac{d\rho(z_1)}{dz}= \infty \; \; ; \; \; \frac{d\rho(z_2)}{dz}= - \infty
\end{equation}
The equation is solved numerically by an iterative procedure checking the
minimization of the deformation energy with a given accuracy.
The phenomenological shell corrections to the LDM deformation energy
presented in section~\ref{binary2} are used to obtain reflection asymmetric
saddle point shapes. In fact
the equation to be solved numerically is obtained from (\ref{eq:sec}) after
changing the variable and function as shown below.

One can develop the computer code for just one of the ``fragments'' (for
example for the right hand one extended from $z=0$ to $z=z_2$) and then
write the result for the other fragment (left hand one from $z=-z_1$ to 
$z=0$). For symmetrical shapes we have $z_2=z_p=-z_1$. It is convenient to
make a change of the function and variable defined by:
\begin{equation}
u(v )={\Lambda}^2\rho^2[z(v )] \; \; ; \; \; z(v )=z_p - v/{\Lambda}
\end{equation}
therefore
$dz/dv=-1/{{\Lambda}}$, $u'=du/dv=2{{\Lambda}}^2\rho (d\rho/dz)dz/dv
=-2{{\Lambda}}\rho \rho '$, $\rho=\sqrt{u}/{{\Lambda}}$, $u'^2=4u\rho '^2$,
$1+\rho'^2=u'^2/(4u)+1$,
$u''=d^2u/dv^2=d(u')/dv=-2{{\Lambda}}[d\rho(z(v))/dv]d\rho/dz
-2{{\Lambda}}\rho d(\rho (z(v))/dz=-2{{\Lambda}}(d\rho/dz)(dz/dv)(d\rho/dz) -
2{{\Lambda}}\rho (d^2\rho/dz^2)dz/dv=2\rho'^2 + 2\rho\rho''$.
By substituting into equation (\ref{eq:sec}) one has
\begin{equation}
u''-2-\frac{1}{u}\left[u'^2+\left(\frac{3XV_s}{2{{\Lambda}}}+\frac{\lambda_1
+\lambda_2z_p}{4{{\Lambda}}} - \frac{\lambda_2v}{4{{\Lambda}}^2}\right)
(4u+u'^2)^{3/2}\right]=0
\end{equation}
A linear function of $v$ is introduced  by adding and subtracting
$a+bv$ to $3XV_s/2{{\Lambda}}$. The quantity $V_{sd}$ is defined as the
deviation of Coulomb potential at
the nuclear surface from a linear function of $v$
\begin{equation}
V_{sd}=\frac{3X}{2{{\Lambda}}}V_s -a -vb
\end{equation}
where the constant  
\begin{equation}
a=\frac{3X}{2{{\Lambda}}}V_s(v=0)
\end{equation}
is chosen to give $V_{sd}(v=0)=0$, and
\begin{equation}
b=\left[\frac{3X}{2{{\Lambda}}}V_s(v=v_p)-a\right]/v_p
\end{equation}
where $v_p=\Lambda z_p$. Consequently one has
\begin{equation}
u''-2-\frac{1}{u}\left\{u'^2+ \left[
\left(\frac{\lambda_1+\lambda_2z_p}{4{{\Lambda}}}+
a \right) +v \left(b -
\frac{\lambda_2}{4{{\Lambda}}^2} \right) + V_{sd} \right](4u+
u'^2)^{3/2}\right\}=0
\end{equation}
By equating with 1 the coefficient of $v$, one can establish
the following link between ${{\Lambda}}$ and the
Lagrange multiplier $\lambda_2$
\begin{equation}
{{\Lambda}}^2=\lambda_2/4(b -1)
\end{equation}
In this way $u(v)$ is to be determined by the equation
\begin{equation}
u''- 2 - \frac{1}{u}[u'^2+(v - d + V_{sd})(4u+u'^2)^{3/2}] = 0
\label{eq:f}
\end{equation}
where the role of a Lagrange multiplier is played by the quantity $d$ which
is taken to be constant instead of $\alpha$. The value
of the deformation coordinate $\alpha$ is calculated after obtaining a
convergent solution. To the tip $z=z_p$, at which $\rho(z_p)=0$, corresponds
$v=0$, hence $u(0)=\Lambda^2\rho^2(z_p)=0$. By multiplying with $u$ the
equation (\ref{eq:f}), introducing $v=0$, and using the relationship
$V_{sd}(v=0)=0$, it follows that $u'(0)=1/d$. Consequently the boundary
conditions for $u(v)$ are:
\begin{equation}
u(0)=0, \; \; u'(0)=1/d
\end{equation}
To $z=0$, at which $\rho'(0)=0$ (the middle of the neck for elongated
shapes), corresponds $v_p=\Lambda z_p$ and
$u'(v_p)=-2\Lambda\rho(0)\rho'(0)=0$. The point $v=v_p$ in which
\begin{equation}
u'(v _{pn})=0
\label{eq:m}
\end{equation}
is determined by interpolation from two consecutive values of
$v_p$ leading to opposite signs of $u'(v)$. The number $n$ of changes of
signs is equal to the number of necks plus one given in advance, e.g. for a
single neck (binary fission) $n=2$ and for two necks (ternary fission)
$n=3$, etc. 

Although the quantity ${{\Lambda}}$ is not present in eq~(\ref{eq:f}) we
have to know it in order to obtain the shape function $u(v)$. By changing
the function and the variable in the eq~(\ref{eq:v}) one has
\begin{equation}
{{\Lambda}} = \left\{ \frac{3}{2} \int_0^{v _{pn}}
u (v )dv \right\}^{1/3}
\label{eq:vv}
\end{equation}
and the deformation coordinate, $\alpha=z^c_L+z^c_R$, may also be
determined by adding to
\begin{equation}
z^{c}_R=2\pi R_0^3\int_{0}^{z_p} z\rho^2(z)dz/V=\frac{3}{2}\int_{v_p}^0
\frac{v_p-v}{\Lambda}\frac{u}{\Lambda ^2}\frac{-dv}{\Lambda} =
\frac{3}{2\Lambda ^4}\int_0^{v_p}(v_{p} - v)u(v)dv
\end{equation}
a similar relationship for $z^c_L$.
From the dependence $\alpha(d)$, one can obtain the inverse function
$d=d(\alpha)$.

In order to find the shape function $u(v)$ we solve eq~(\ref{eq:f}) with
boundary conditions written above. One
starts with given values of the constants $d$ and $n$. For reflection
symmetric shapes $d_L=d_R$ and $n_L=n_R$. 
In the first iteration one obtains
the solution for a Coulomb potential at the nuclear surface assumed to be a
linear function of $v$, i.e. for $V_s=0$. Then one calculates the parameters
${{\Lambda}}$, $a$, and $b$, which depend on the
Coulomb potential and its deviation $V_{sd}$ from a linear function, and the
deformation energy corresponding to the nuclear shape
\cite{p75cpc78,p80cpc80}. The
quantity $V_{sd}$ determined in such a way is introduced in eq~(\ref{eq:f})
and the whole procedure is repeated until the deformation energy is obtained
with the desired accuracy. In every iteration the equation is solved
numerically with the Runge-Kutta method.

One can calculate for different values of deformation $\alpha$ (in fact for a
given $d_L$ and $d_R$) the deformation energy $E_{def}(\alpha)$. The
particular value $\alpha_s$ for which $dE_{def}(\alpha _s)/d\alpha=0$
corresponds to the extremum,  i.e. the shape function describes
the saddle point, and the
unconditional extremum of the energy is the fission barrier.  
The other surfaces (for $\alpha \neq \alpha_s$) are extrema only with 
condition $\alpha=$~constant. 
In this way one can compute the deformation energy versus $d_L=d_R$.
In Fig.~\ref{fig5}
one can see an example of variation of deformation parameter and the
deformation energy with $d_L$ for $^{238}$U at symmetry
$\eta=0$. The saddle point corresponds to the maximum of deformation energy.

For {\em reflection asymmetrical shapes}
we need to introduce another constraint: the asymmetry parameter,
$\eta$, defined by
\begin{equation}
\eta=\frac{M_L-M_R}{M_L+M_R} = \frac{A_1 - A_2}{A_1 + A_2}
\end{equation}
It should remain constant during  variation of
the shape function $u(v)$.
Consequently eq~(\ref{eq:f}) should be written differently for left hand
side and right hand side.
Now $d_L$ is different from $d_R$, and so are the
parameters ${{\Lambda}}_L$ and ${{\Lambda}}_R$. They have to fulfil matching
conditions
\begin{equation}
\rho_L(z=0)=\rho_R(z=0)
\end{equation}
hence
\begin{equation}
u_L^{1/2}(v_p)/{{\Lambda}}_L = u_R^{1/2}(v_p)/{{\Lambda}}_R
\end{equation}
The similar condition for derivatives $\rho'(z)$ in $z=0$,
\begin{equation}
\rho'_L(z=0)=\rho'_R(z=0)=0
\end{equation}
is automatically satisfied due to eq~(\ref{eq:m}).
The second derivative $\rho''(z)$  can have a discontinuity in $z=0$
if $d_L \neq d_R$.
The parameters ${{\Lambda}}_L$ and ${{\Lambda}}_R$ 
are easily expressed in terms of $\eta$, if we write the definition of mass
asymmetry as
\begin{equation}
M_L=\frac{2\pi}{3}(1+\eta)=\pi {{\Lambda}}_L^{-3}\int_0^{v_p}u_L(v)dv
\end{equation}
\begin{equation}
M_R=\frac{2\pi}{3}(1-\eta)=\pi {{\Lambda}}_R^{-3}\int_0^{v_p}u_R(v)dv
\end{equation}
We assume that $M_L+M_R$ is equal to the mass of a sphere with $R_0=1$.
It follows
\begin{equation}
{{\Lambda}}_L=(1+\eta)^{-1/3}{{\Lambda}}_{L0}
\end{equation}
\begin{equation}
{{\Lambda}}_R=(1-\eta)^{-1/3}{{\Lambda}}_{R0}
\end{equation}
where we introduced notations similar to eq~(\ref{eq:vv}):
\begin{equation}
{{\Lambda}}_{L0(R0)}=\left\{ \frac{3}{2}\int_0^{v_p}u_{L(R)}(v)dv
\right \}^{1/3}
\end{equation}
The shape  of a nucleus with given mass asymmetry, $\eta$, is completely
determined by the above written equations in which the quantities $u_L(v_p)$
and $u_R(v_p)$ are solutions of the eq~(\ref{eq:f}). There is an almost
linear dependence of $\eta$ from the difference $d_L-d_R$.

\section{Mass symmetry in binary fission within LDM}\label{binary}

One can test the method by comparing some nuclear shapes within LDM to the
standard results for medium and heavy nuclei. In Figure~\ref{fig6} we
present reflection symmetric nuclear shapes for binary fission of a nucleus
with the fissility parameter $X=0.6$ (e.g. $^{170}$Yb), obtained for
$n_L=n_R=2$ (one neck), $d_L=d_R=1.4; 1.5; 1.7$, and $1.91$ (for which
$\alpha=1.314; 1.644; 2.100$ and $2.304$) and a vanishing mass asymmetry
$\eta=0$.  The saddle point (maximum value of the conditioned deformation
energy minimum) is obtained for $d_L=1.91$, at which the shape is deformed
and necked-in.

A comparison between three nuclear shapes at the saddle point 
for nuclei with fissilities $X= 0.60, 0.70,$ and $0.82$ (corresponding to 
$^{170}$Yb, $^{204}$Pb, and  $^{252}$Cf nuclei lying on the line 
of beta-stability)
is presented in Figure~\ref{fig9}. One can see how the necking-in 
and the elongation are decreasing 
($\alpha = 2.304; 1.822$ and $1.165 $) when fissility increases from
$X=0.60$ to $X=0.82$, in agreement with \cite{coh63ap}.
In the limit $X=1$ the saddle point shape is spherical.
The method proved its capability by reproducing the well known LDM saddle
point shapes.

\section{Mass asymmetry in binary fission}\label{binary2}
\begin{figure}[ht]
\centerline{\includegraphics[width=7.5cm]{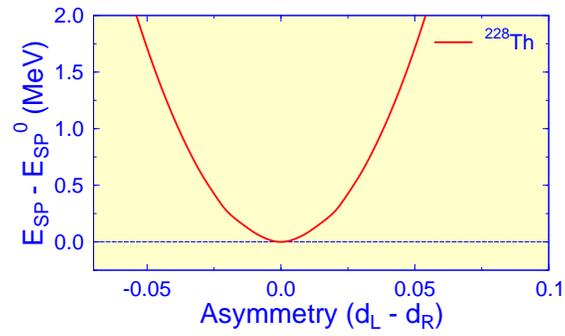}}
 \caption{The saddle point deformation energy $E_{SP}$ vs. mass-asymmetry 
within a pure liquid drop model. Example of $^{228}$Th.\label{fig13}}
 \end{figure}

\begin{figure}[hb]
\centerline{\includegraphics[width=6.cm]{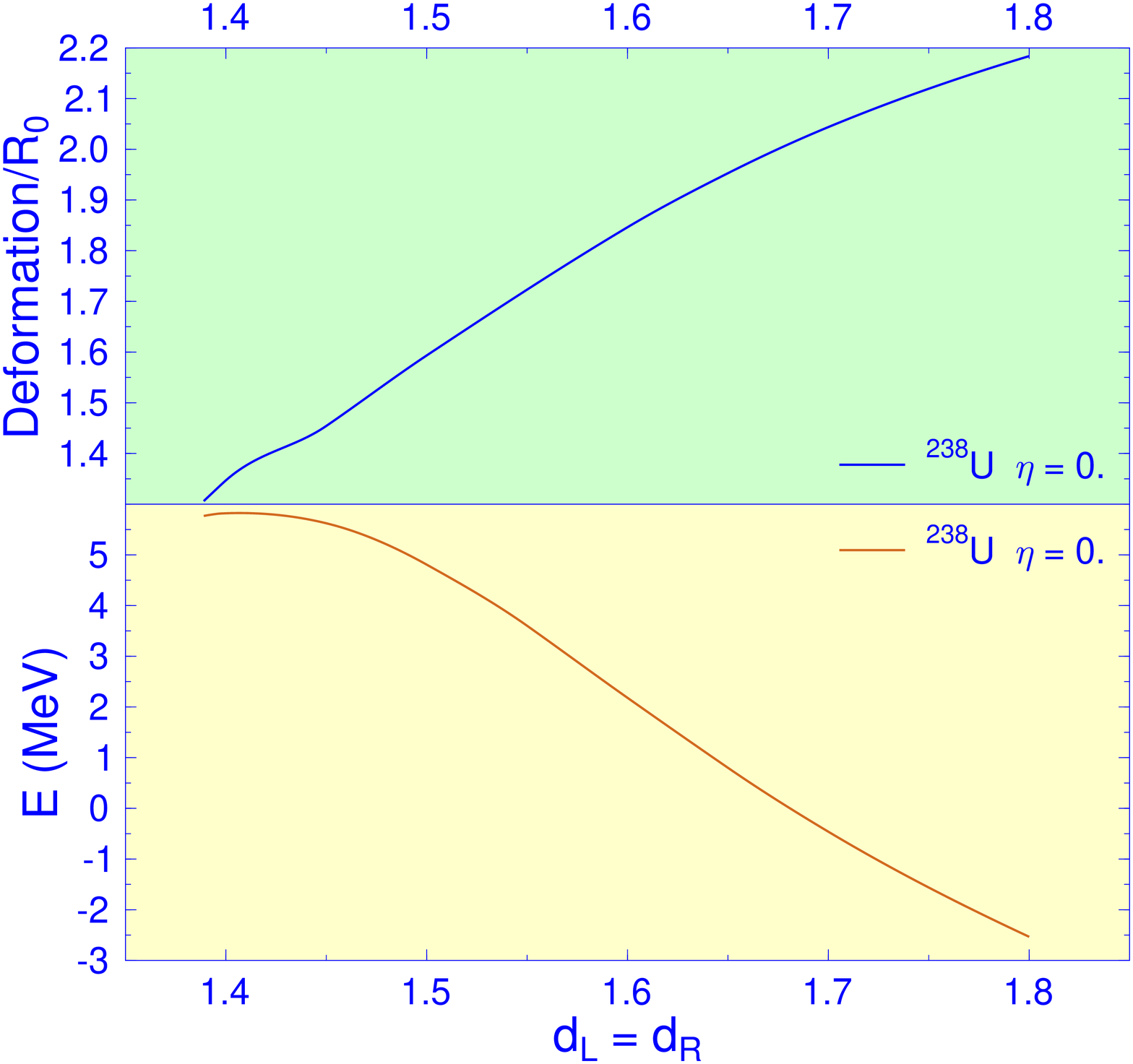}}
 \caption{Deformation and potential energy vs. input parameter of the
integro-differential equation for binary fission of $^{238}$U within a pure
LDM. 
\label{fig5}}
 \end{figure}

\begin{figure}[hb]
\centerline{\includegraphics[width=7.5cm]{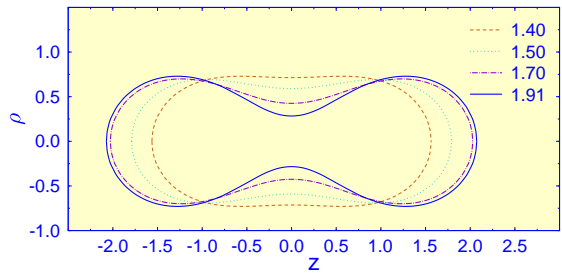}}
 \caption{Nuclear shapes during binary fission of a nucleus with fissility
$X=0.60$ for $d_L=d_R=1.40, 1.50, 1.70, 1.91$.
\label{fig6}}
 \end{figure}

\begin{figure}[ht]
\centerline{\includegraphics[width=7.cm]{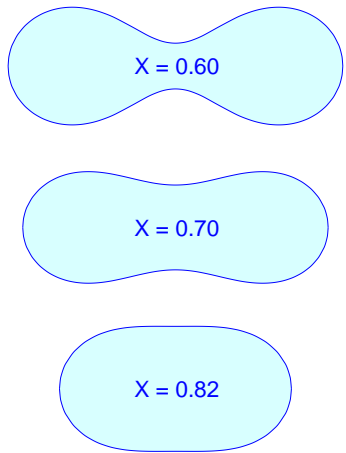}}
 \caption{Sadle-point shapes during binary fission of nuclei with fissility
$X=0.60, 0.70, 0.82$. 
\label{fig9}}
 \end{figure}

\begin{figure}[hb]
\centerline{\includegraphics[width=8cm]{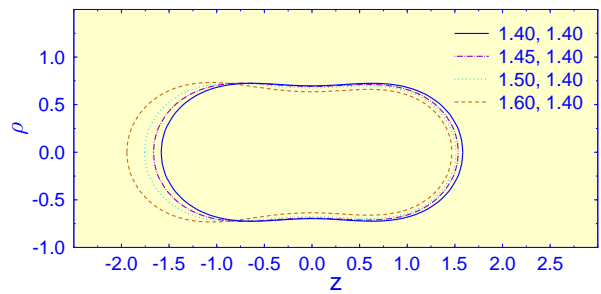}}
 \caption{Mass asymmetric shapes during binary fission of a nucleus with fissility
$X=0.60$ for $d_L \ne d_R$. Pure LDM calculations. \label{fig8}}
 \end{figure}

Within LDM a nonzero mass asymmetry parameter
(see the shapes from figure~\ref{fig8}) leads to a deformation energy which
increases with $\eta$ as is illustrated in~figure~\ref{fig13}, 
where $\eta$ is replaced by an 
almost linear dependent quantity $(d_L-d_R)$.
The reflection asymmetric shapes plotted in figure~\ref{fig8}, resulted by
choosing the input parameters as follows: $n_L=n_R=2$; $d_L=1.40; 1.45;
1.50$, and $1.60$ while $d_R=1.40$ was kept constant, and so was $X=0.60$.
The increasing deformation energy with mass-asymmetry in figure~\ref{fig13},
refers to different values of fissility, namely 
$X=0.758$ for $^{228}$Th.

When the shell effects are taken into account a saddle point solution of the
integro-differential equation with reflection asymmetry is obtained. In the
following we shall use a phenomenological shell correction adapted after
Myers and Swiatecki \cite{mye66np}. At a given deformation one calculates
the volumes of fragments and the corresponding numbers of nucleons
$Z_i(\alpha), \ N_i(\alpha)$ ($i=1,2$), proportional to the volume of each
fragment. Then one can add for each fragment the contribution of protons and
neutrons
\begin{equation}
\delta E(\alpha) = \sum_i \delta E_i(\alpha) = \sum_i [\delta E_{pi}(\alpha) 
+ \delta E_{ni}(\alpha)] 
\end{equation}
given by
\begin{equation}
 \delta E_{pi}=Cs(Z_i) ; \ \ \delta E_{ni}=Cs(N_i)
\end{equation}
where
\begin{equation}
 s(Z) = Z^{-2/3}F(Z) -cZ^{1/3} 
\end{equation}
and similar eq for $s(N)$.
\begin{equation}
F(n) = \frac{3}{5}\left [\frac{N_i^{5/3} -N_{i-1}^{5/3}}{N_i -
N_{i-1}}(n -N_{i-1}) - n^{5/3}+ N_{i-1}^{5/3} \right ]
\end{equation}
where $n \in (N_{i-1}, N_i)$ is the current number of protons ($Z$) or
neutrons ($N$) and $N_{i-1}, N_i$ are the nearest magic numbers. The
parameters $c=0.2$, $C=6.2$~MeV were determined by fit to experimental masses
and deformations.

The dependence on deformation~\cite{sch71pl} $\alpha$ is given by
\begin{equation}
\delta E(\alpha) = \frac{C}{2}\left \{ \sum_i[s(N_i)+
s(Z_i)]\frac{L_i(\alpha)}{R_i} \right \}
\end{equation}
where $L_i(\alpha)$ are the lenghths of fragments along the symmetry axis.
During the deformation process, the variation of separation distance between 
centers, $\alpha$, induces the variation of the geometrical quantities and of
the corresponding nucleon numbers. Each time a proton or neutron number reaches
a magic value, the correction energy passes through a minimum, and it has a
maximum at midshell.

\begin{figure}[hb]
\centerline{\includegraphics[width=8cm]{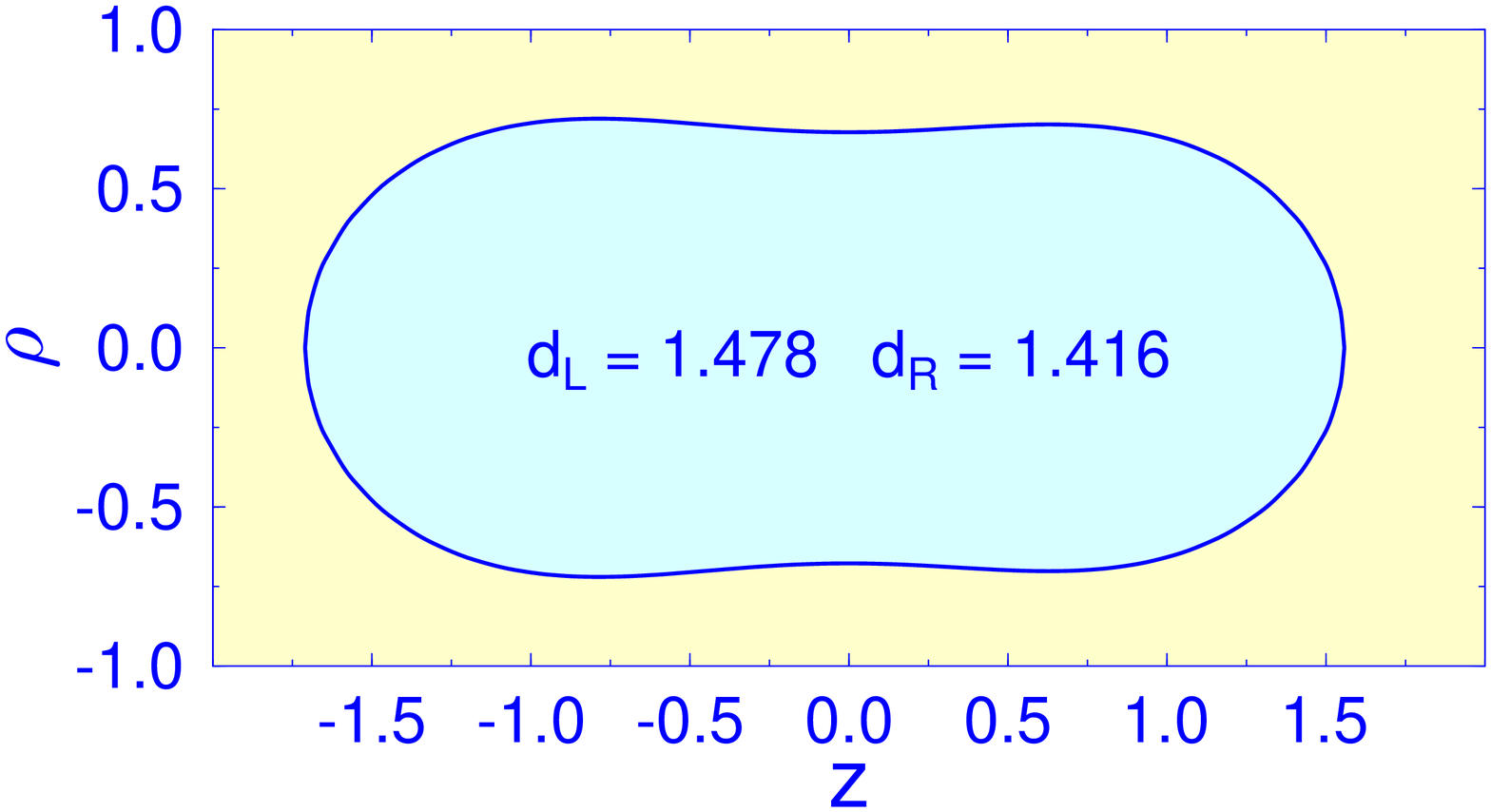}}
 \caption{Mass asymmetric saddle point shape of $^{232}$U.  Shell effects
taken into account. \label{fig4}}
 \end{figure}

\begin{figure}[htb]
\centerline{\includegraphics[width=8.5cm]{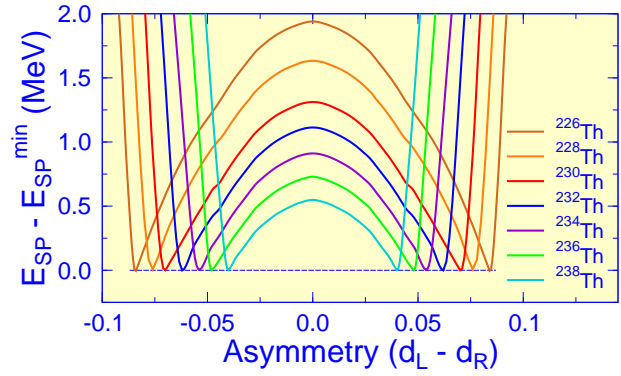}}
 \caption{Difference between the saddle point deformation energy $E_{SP}$
and its minimum value $E_{SP}^{min}$ vs mass asymmetry parameter
[related to $(d_L-d_R)$] for binary fission of Th isotopes.
One can see the minima produced by the shell effects.\label{Fig.9}}
 \end{figure}

\begin{figure}[htb]
\centerline{\includegraphics[width=8.5cm]{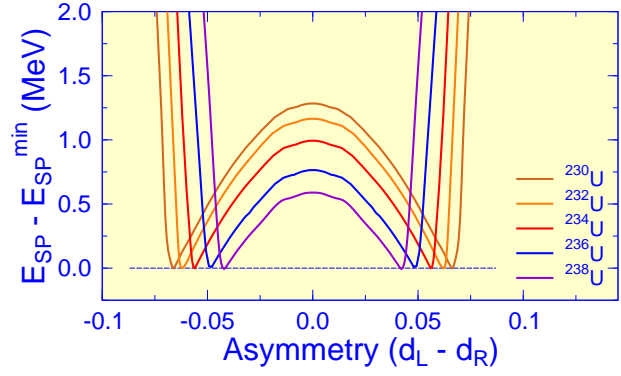}}
 \caption{Difference between the saddle point deformation energy $E_{SP}$
and its minimum value $E_{SP}^{min}$ vs mass asymmetry parameter
[related to $(d_L-d_R)$] for binary fission of U isotopes
in the presence of shell corrections.\label{Fig.10}}
 \end{figure}

Results for binary cold fission of parent nuclei $^{226-238}$Th and
$^{230-238}$U are presented in figures~\ref{Fig.9} and \ref{Fig.10}.  The
minima of the saddle point energy occur at nonzero mass asymmetry parameters
$d_L-d_R$ in the range $0.04, 0.08$ for the above mentioned nuclei. They
correspond to $\eta$ of $0.050, 0.095$ which leads to $A_1 \simeq 125$ in
all cases. 
A typical saddle point
shape, for $^{232}$U may be seen in figure~\ref{fig4}. For experimentally
determined mass asymmetry \cite{gun69ar,cro77adndt} 
the maximum of the fission fragment mass distributions is centered on
$A_1=140$ in a broad range of mass numbers of parent nuclei.

\begin{figure}[htb]
\centerline{\includegraphics[width=6cm]{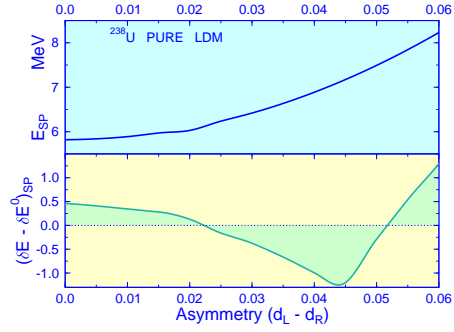}} 
 \caption{The saddle point deformation energy $E_{SP}$ of $^{238}$U
within a pure liquid drop model (top). 
The minimum of the $E_{SP}$ is produced 
by the negative values of the shell corrections $(\delta E - \delta
E^0)_{SP}$ (bottom).\label{Fig.4}}
\end{figure}

In order to understand correctly the figure~\ref{Fig.9}, where from 
the saddle point energies $E_{SP}$ of every nucleus we subtracted its
minimum value $E_{SP}^{min}$,
we would like to give an example for $^{238}$U in figure~\ref{Fig.4}. In the
upper part we plot the saddle point energies obtained within a pure LDM (see
also Fig.~\ref{fig13}).
When we add the shell corrections, the conditions of equilibrium are changed
and in general the LDM part of the saddle point energy is not identical with
the previous one, as may be also seen from the tables~\ref{table3} 
and~\ref{table4}. 
Table ~\ref{table4} shows how are changed equilibrium conditions when the
shell effects are taken into account. There are two rows for every value of
$d_L-d_R$ when $d_L-d_R < 0.045$: one for maximum value of the total
$E_{SP}$, and the other one for maximum value of its LDM term $E_{SP-LDM}$.

The minimum of the $E_{SP}$  is produced by the negative
values of the shell 
corrections $\delta E - \delta E^0$ which can be
clearly seen in the lower part of the figure~\ref{Fig.4}. 
The variation of the saddle point energy with the mass asymmetry parameter
$d_L - d_R$ is almost a linear function of the mass asymmetry $\eta$
for some even-mass isotopes of Th and U. The
minima of the saddle point energy occur at nonzero mass asymmetry parameters
$d_L-d_R$ between about 0.04 and 0.085 for these nuclei. When the mass
number of an isotope increases, the value of the mass asymmetry
corresponding to the minimum of the SP energy decreases.

As mentioned by Wilkins et al. \cite{wil76pr}, calculations of PES for
fissioning nuclei ``qualitatively account for an asymmetric division of
mass''. From the qualitative point of view the results displayed in
Figures~\ref{Fig.9} and \ref{Fig.10}
proove the capability of the method to deal with fission mass and
charge asymmetry. 
The experimentally determined mass number of the most probable heavy
fragment \cite{wah88adnd} for U isotopes ranges from 134 to 140.  The
corresponding values at the displayed minima in Figures~\ref{Fig.9} and
\ref{Fig.10} are very
close to 125, which means a discrepancy between 6.7~\% and 10.7~\% for
$A_H$. 
The inaccuracy in reproducing the experimental mass asymmetry is due to the
contribution of the phenomenological shell corrections. In the absence of
shell corrections the pure liquid drop model (LDM) reflection-symmetric
saddle point shapes \cite{coh63ap} are reproduced, and the barrier height
increseas with an increased mass asymmetry. When the shell corrections are
taken into account the LDM part behaves in the same manner (larger values at
non-zero mass asymmetry). Only the contribution of shell effects can produce
a minimum of the barrier height at a finite value of the mass asymmetry. One
may hope to obtain a better agreement with experimental data by using a more
realistic shell correction model, based for example on the recently
developed two center shell model \cite{rad03prc}.

\section{Ternary Fission}\label{ternary}
Neutron multiplicities higher than one, in the induced nuclear
fission, are used to produce the chain reaction, on which the nuclear
energetics is based. The condition of a positive released energy, $Q > 0$, 
in such a complex process is easily fulfilled, and the escape of one
or several neutrons from the neck formed between
the light- and heavy fragment, is not prevented by any Coulomb barrier.
A small and narrow centrifugal barrier, due to the angular momenta carried
away by the neutrons, do not constitute a major obstacle.
A charged particle has to penetrate, by quantum tunneling, a much thicker 
and higher potential
barrier, leading to a long delay and to a corresponding comparable low yield.
Nevertheless, the particle-accompanied fission (or ternary fission) was
observed both in neutron-induced and spontaneous fission since 1946.
Several such processes, in which the charged particle is a proton, deuteron, 
triton, $^{3-8}$He, $^{6-11}$Li,
$^{7-14}$Be, $^{10-17}$B, $^{13-2}$C, $^{15-20}$N,   $^{15-22}$O,
have been detected \cite{mut96mb}. Many other heavier isotopes of F, Ne, Na,
Mg, Al, Si, P, S, Cl, Ar, and even Ca were also mentioned.

\begin{figure}[htb]
\centerline{\includegraphics[width=8cm]{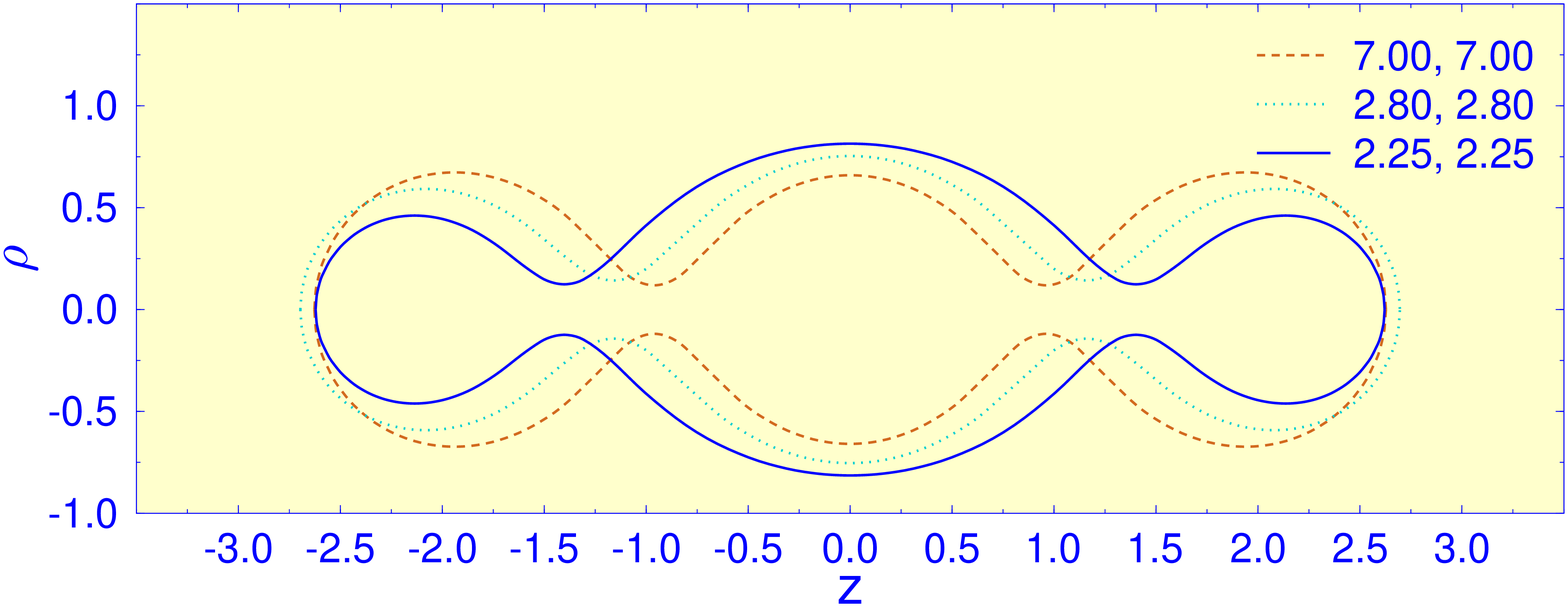}}
\caption{Shapes obtained by solving an integro-differential equation 
for $n_L= n_R=3$, $d_L=d_R=2.25, 2.80,$~and
7.00. The binary fissility $X=0.60$ corresponds to $^{170}$Yb.\label{fig2}}
\end{figure}

Different elongated shapes for ternary fissions are shown in
figure~\ref{fig2}. 
For shapes with three fragments and two
necks ($n_L= n_R=3$) when $d_L=d_R$ is increased 
from 2.25 to 2.80 and 7.00 the deformation $\alpha$
increases from 1.650 to 2.306 and 2.730. In the same time the elongation
is initially increased from 5.234 to 5.392 and then decreased to 5.24;
the fragment radii are 0.461/0.814/0.461, 0.592/0.753/0.592, and
0.673/0.659/0.673, leading to decreasing energies in units of $E_s^0$
from 0.165 to 0.150 and 0.134.
The configuration with $E/E_s^0=0.134$ is not far from a 
{\em ``true ternary-fission''} in which the 
three fragments are almost identical:
$^{170}_{70}$Yb$\rightarrow ^{56}_{23}$V + $^{56}_{23}$V + $^{58}_{24}$Cr
and the $Q$-value is 83.639~MeV.
One may compare the above $E/E_s^0$ value with the touching-point energy
of these spherical fragments $(E_t - Q)/E_s^0 = 0.239$. It is larger, as
expected, because of the finite neck of the shapes in figure~\ref{fig2}. 
For $\alpha$-accompanied fission of $^{170}$Yb with two
$^{83}_{34}$Se fragments $Q=87.484$~MeV is larger and the touching point
energy $(E_t - Q)/E_s^0 = 0.103$ is lower. A lower $Q=70.859$~MeV and higher
energy barrier $(E_t - Q)/E_s^0 = 0.147$ is obtained for $^{10}$Be
accompanied fission of $^{170}$Yb with $^{80}_{33}$As fission fragments.

\begin{figure}[htb]
\centerline{\includegraphics[width=4cm]{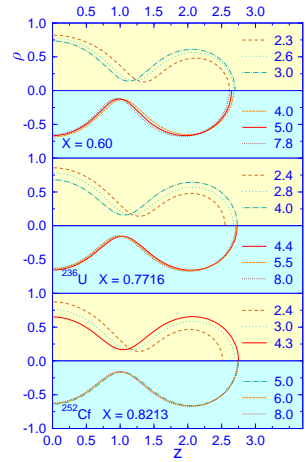}} 
 \caption{Evolution of ternary shapes ($n_L= n_R=3$) when the input 
paramter $d_L=d_R$ is increased as shown for $^{170}$Yb~(top), 
$^{236}$U~(middle), and $^{252}$Cf~(bottom). The total length
on the symmetry axis increases with increasing $d_L=d_R$ in the upper part
of the panel and decreases in the lower part.\label{Fig.5}}
 \end{figure}

\begin{figure}[htb]
\centerline{\includegraphics[width=7.5cm]{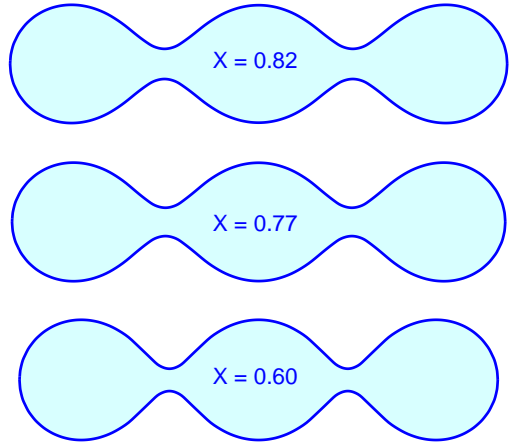}}
 \caption{Ternary shapes ($n_L= n_R=3$) approaching a scission into three
fragments with identical radii $^{170}$Yb~(bottom), $^{236}$U~(middle), 
and $^{252}$Cf~(top).\label{Fig.3}}
 \end{figure}

Figure~\ref{Fig.5} shows how evolves the ternary shapes for three values of
the fissility ($X=0.60$ corresponding to $^{170}$Yb, $X=0.7716$ corresponding
to $^{236}$U and $X=0.8213$ e.g. $^{252}$Cf) when  the input
paramter $d_L=d_R$ is increased. For smaller values of $d_L=d_R$ the
elongation of the shapes becomes larger up to a limiting value of about 3, 4
and 4.3 respectively; then if $d_L=d_R$ is further increased it becomes
slightly smaller. 
In what concerns the shapes approaching the scission into three identical
fragments (figure~\ref{Fig.3}) the total length increases with increasing
fissility.

Systematic calculations~\cite{p209adndt98} have shown a clear
correlation between the $Q$-values and the measured yield of different
isotopes for one cluster accompanied fission. For example, among the He
isotopes with mass numbers 4,~6,~and~8, $^4$He leads to the maximum 
$Q$-value.
The maximum yield was indeed  experimentally observed~\cite{mut96mb} for
$\alpha$~accompanied fission. Similarly, among
$^{6,8,10,12}$Be, the clusters
$^8$Be and $^{10}$Be give the maximum $Q$-values. 
As $^8$Be spontaneously
breaks into 2$\alpha$ particles it is not easy to measure $^8$Be accompanied
fission yield; consequently $^{10}$Be has been most frequently
identified. {\em By detecting, in coincidence, these two alpha particles,
the $^8$Be accompanied fission
with a larger yield compared to that of the $^{10}$Be one, could be observed 
in the future}.
From $^{12,14,16,18}$C the favoured is $^{14}$C, and all
$^{16,18,20,22}$O isotopes have comparable $Q$-values when they are emitted
in a cold binary fission of $^{252}$Cf. Nevertheless, $^{20}$O is slightly
upper than the others.
As a rule, if the $Q$-value is larger the barrier height is smaller, and the
quantum tunneling becomes more probable. The stronger emission of
$^{14}$C compared to $^{12}$C has the same explanation as for the
$^{14}$C radioactivity; the $Q$-value is larger
because the heavy fragment is doubly magic.

We should stress again that if one is interested to estimate the yield 
in various fission processes, one has to compare 
the potential barriers and not the $Q$-values. Our results are in agreement
with preceding calculations \cite{die74np} showing also preference for
prolate over oblate shapes.
Theoretically it was pointed out by Present \cite{pre41pr} in 1941 that
Uranium tripartition would release about 20~MeV more energy than the binary
one. In spite of having quite 
large Q values \cite{p209adndt98}, this ``true ternary fission'' is a 
rather weak process; the strongest phenomenon remains the 
$\alpha $-particle-accompanied fission.

Experiments on so-called ``symmetrical tripartition'' were performed {\em
e.g.} using the induced fission of $^{235}$U by thermal neutrons 
\cite{ros50pr},
the induced fission of $^{238}$U by intermediate-energy hellium ions
\cite{iye68pr}, or spontaneous fission of $^{252}$Cf \cite{sch87pl}, etc.
An yield of $6.7 \pm 3.0$ per $10^6$ binary fissions was reported by Rosen
and Hudson \cite{ros50pr} who employed a triple gas-filled
ionization chamber and a suitable electronics including a triple coincidence
circuit. Other ``optimistic'' results are mentioned by Iyer and Cobble
\cite{iye68pr}, who tried radiochemical methods of identification at
intermediate energy of excitation. While at high energy \cite{fle66pr}
implying bombarding heavy ions of several hundred MeV, a positive result may
be accepted, it is not certain whether it comes out from a compund
nucleus. The general conclusion \cite{sch87pl} after measuring triple
coincidences with detectors placed at 120$^{\circ}$ is 
rather pessimistic: except 
for excitation energies over 24~MeV, the true ternary yield is extremely
low: under 10$^{-8}$ per binary fission act.
By performing dynamical calculations, Hill
arrived in his thesis and in \cite{hil58c} at elongated shapes with
pronounced necks looking more encouraging for particle-accompanied fission. 
It would be rewarding to perform successful experiments with nowadays very
much improved experimental techniques,
despite the previous rather pessimistic 
conclusion that ``true'' ternary spontaneous fission is an extremely rare 
phenomenon.

\section{Multicluster fission }\label{multi}  
The shapes with four fragments and three necks ($n_L= n_R=4$) can be seen in
figure~\ref{fig10}. When $d_L=d_R$ increases from 2.30 to  2.70 and 4.00  
the deformation takes the values 2.144, 3.136, and 3.233 and the elongations
are 6.077, 6.916 and 6.252.
The fragment radii are 0.408/0.625/0.625/0.408, 0.479/0.632/0.632/0.479, and
0.608/0.616/0.616/0.608, for which the energies in units of $E_s^0$  are
0.188, 0.216 and 0.214, respectively.
The last shape,  with $E/E_s^0=0.214$ approaches a fission into
almost identical four fragments $^{170}_{70}$Yb$\rightarrow ^{42}_{17}$Cl +
$^{42}_{17}$Cl + $^{43}_{18}$Ar + $^{43}_{18}$Ar. Again the configuration
with aligned spherical fragments in touch is higher in energy: 
$(E_t - Q)/E_s^0 = 0.324$.
Even more complex shapes can be obtained by further increasing the values of 
$n_L= n_R$.

\begin{figure}[htb]
\centerline{\includegraphics[width=8cm]{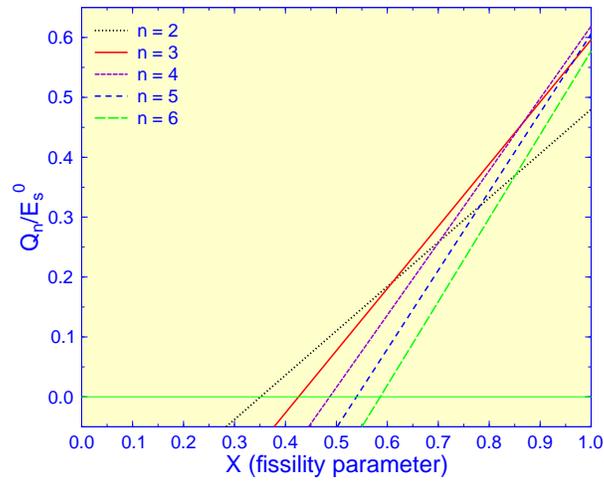}}
 \caption{Approximation of $Q$-values for fission into equally sized
fagments.
\label{fig3}}
 \end{figure}

\begin{figure}[ht]
\centerline{\includegraphics[width=7cm]{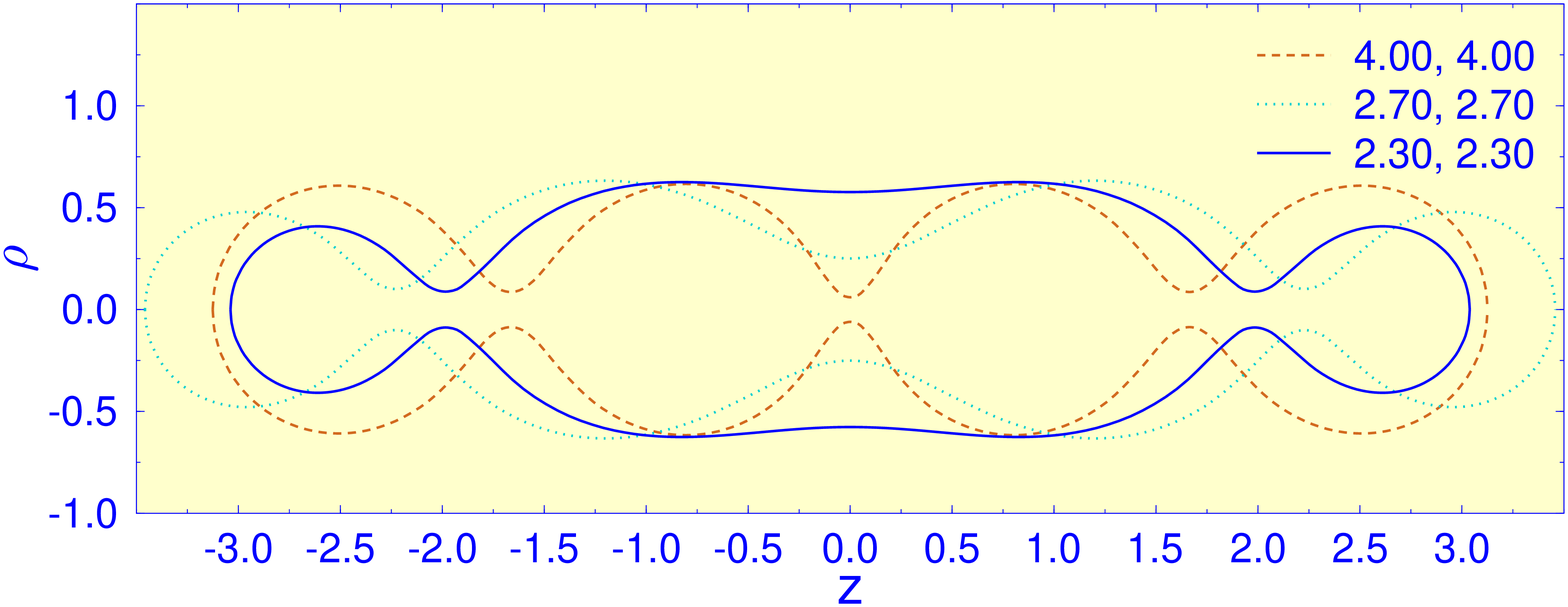}}
 \caption{Nuclear shapes during quaternary fission of a nucleus with fissility
$X=0.60$ for $d_L=d_R=2.30, 2.70, 4.00$.
\label{fig10}}
 \end{figure}

In 1958 it was theoretically shown \cite{swi58c}
on the basis of the liquid drop model \cite{mye66np}
that for increasingly heavier nuclei, fission into three, then four
and even five fragments becomes energetically more favourable than binary
fission (see figure~\ref{fig3}). 
One can take, as an approximation of the $Q$ value, 
the energy difference between the sum of Coulomb and surface energies    
for the parent (superscript~$^0$) and $n$ identical fission fragments    
(superscript~$^i$)
\begin{equation}  
Q_n \simeq (E^0_C + E^0_s) - \sum_{i=1}^n(E^i_C + E^i_s)
\end{equation}
where $n=2$ for binary fission, $n=3$ for ternary fission, 
$n=4$ --- a quaternary fission, $n=5$ --- a division into
five fragments, $n=6$ --- a division into six fragments, etc.

A linear
dependence of $Q_n$ on the (binary) fissility parameter, 
$X=E^0_C/(2E^0_s)$, of the form 
\begin{equation}
Q_n/E^0_s \simeq 1-n^{1/3} + 2X(1-n^{-2/3})
\end{equation}
has been obtained \cite{swi58c}.
When the fissility parameter increases, fission into
more than two equal fragments becomes energetically favored. At $X \geq
0.426$ tripartition becomes exothermic and for  $X \geq 0.611$ the $Q$-value
for fission into three identical fragments is larger than that for binary
fission.
The general trend, and sometimes even the absolute
values of $Q_2$ and $Q_3$, are well reproduced  \cite{p209adndt98}
by the above equation.

A better chance to be experimentally observed has a quaternary
fission in which two light particles are emitted from a neck formed between
two heavy fragments \cite{p218pr99,p231jpg01}. The successful experiment
\cite{gon02pc,gon03hip} on $2\alpha$-accompanied fission observed in cold
neutron induced fission of $^{233,235}$U, confirmed our expectations.

The possibility of a whole family of new decay modes, {\em the multicluster 
accompanied fission}, was envisaged
\cite{p215jpg99,p218pr99,p220pc99,p231jpg01}.
Besides the fission into two or three fragments, a heavy or 
superheavy nucleus spontaneously breaks into four, five or six
nuclei of which two are asymmetric or symmetric heavy fragments and the
others are light clusters, e.g. $\alpha$-particles, $^{10}$Be, $^{14}$C, 
$^{20}$O, or combinations of them.
Examples were presented for the two-, three- and four cluster accompanied
cold fission of $^{252}$Cf and $^{262}$Rf, in which the emitted clusters 
are:  2$\alpha$, $\alpha +
^6$He, $\alpha + ^{10}$Be, $\alpha + ^{14}$C, 3$\alpha$, $\alpha + 
^6$He~+~$^{10}$Be, 2$\alpha + ^6$He, 2$\alpha + ^8$Be, 2$\alpha + ^{14}$C, 
and 4$\alpha$.
A comparison was made with the recently observed $^{252}$Cf
cold binary fission, and cold ternary (accompanied by $\alpha$ particle
or by $^{10}$Be cluster). 
The strong shell effect corresponding to
the doubly magic heavy fragment $^{132}$Sn is emphasized.
From the analysis of different configurations of fragments in touch,
we conclude that the most favorable mechanism of such a decay mode
should be the cluster emission from an elongated neck formed between the two
heavy fragments.

In a first approximation, one can obtain an order of magnitude of the
potential barrier height by assuming spherical shapes of all the
participant nuclei. This assumption is realistic if the fragments are magic
nuclei. For deformed fragments it leads to an overestimation of the barrier.
By taking into account the prolate deformations, one can get smaller
potential barrier height, hence better condition for multicluster emission.
We use the
Yukawa-plus-exponential~(Y+E) double folded model~\cite{sch69zp,kra79pr}
extended~\cite{p80cpc80} for different charge densities. 
In the decay process from one parent to several fragments, the nucleus 
deforms, reaches the touching
configuration, and finally the fragments became completely separated.

Within the Myers-Swiatecki's liquid drop model
there is no contribution of the surface energy to
the interaction of the separated fragments; the deformation energy has
a maximum at the touching point configuration.
The proximity forces acting at small separation distances (within the
range of strong interactions) give rise in the Y+EM to a
term expressed as folllows
\begin{equation}
E_{Yij} = -4\left ( \frac{a}{r_0} \right ) ^2 \sqrt {a_{2i} a_{2j}}
\left [ g_i g_j \left ( 4+\frac{R_{ij}}{a} \right ) 
-g_jf_i - g_if_j \right ] \frac{\exp (- R_{ij}/a)}{R_{ij}/a}
\end{equation}
where
\begin{equation}
g_k =  \frac{R_k}{a} \cosh \left ( \frac{R_k}{a}  \right ) - \sinh
\left ( \frac{R_k}{a} \right )
\; ; \; f_k = \left (\frac{R_k}{a}
\right ) ^2 \sinh \left ( \frac{R_k}{a} \right )
\end{equation}
in which $R_k$ is the radius of the nucleus $A_kZ_k$, 
$a=0.68$ is the diffusivity parameter,
and $a_{2i}$, $a_{2j}$ are expressed in terms of the model
constants $a_s$, $\kappa$ and the nuclear composition parameters
$I_i$ and $I_j$, $a_2 =
a_s(1-\kappa I^2)$, $a_s=21.13$~MeV, $\kappa = 2.3$, $I=(N-Z)/A$,
$R_0=r_0A^{1/3}$, $r_0=1.16$~fm is the radius constant, and $e$ is the 
electron charge,  $e^2 \simeq 1.44$~MeV$\cdot$fm.

\begin{figure}[hb]
\centerline{\includegraphics[width=6cm]{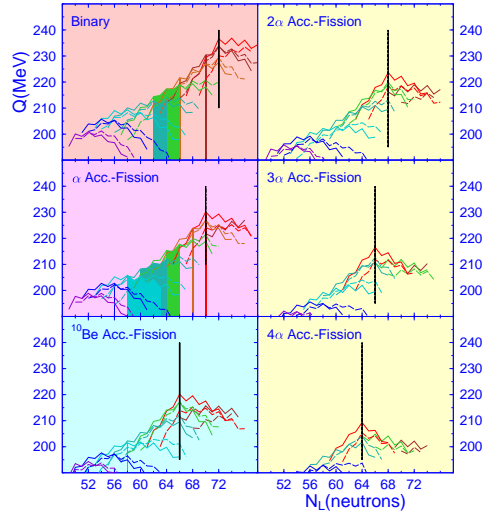}}
 \caption{
Q-values for the cold fission of $^{252}$Cf vs.
neutron number of the light fragment. The experimentally determined
cold fission, $\alpha$-, and $^{10}$Be accompanied fission
on the left hand side are emphasized.
The vertical heavy
bar on each graph corresponds to a magic neutron number  of the
heavy fragment $N_H=82$. 
\label{fig12}}
 \end{figure}

In order to emphasize the strong shell effect, we have chosen in the abscisa 
of the figure~\ref{fig12}
the number of neutrons of the light fragment, $N_L$, which in turn
corresponds to a well defined neutron number of the heavy fragment, $N_H$, 
for a given parent and a given cluster accompanied fission, in our case to: 
$N_H=150-N_L$ for 2$\alpha$ accompanied fission, $N_H=148-N_L$ for 3$\alpha$
accompanied fission, and $N_H=146-N_L$ for 4$\alpha$ accompanied
fission of $^{252}$Cf.
The vertical heavy bar on each plot helps to determine the position of
the magic number $N_H=82$. In a similar way, different types of lines are
drawn through the points belonging to the
same combination of atomic numbers of the fragments $Z_L - Z_H$. 
The red full line is always
reserved for the pair in which the heavy fragment is an isotope of Sn, (with
$Z_H=50$ magic number). The conclusion is clear: the maximum Q-value is
obtained when the heavy fragment is the doubly magic $^{132}$Sn.

In figure~\ref{fig12} the investigated pairs are emphasized. These 
are~\cite{ham94jpg} for the binary fission:
$^{102,104}_{40}$Zr--$^{150,148}_{58}$Ce ($N_L=62,64$), 
$^{104-108}_{42}$Mo--$^{148-144}_{56}$Ba ($N_L=62-66$),  
$^{110}_{44}$Ru--$^{142}_{54}$Xe ($N_L=66$),  and 
$^{116}_{46}$Pd--$^{136}_{52}$Te ($N_L=70$).
For cold $\alpha $ accompanied fission~\cite{ram98prl} one has:
$^{92}_{36}$Kr--$^{156}_{60}$Nd ($N_L=56$), 
$^{96-101}_{38}$Sr--$^{152-147}_{58}$Ce ($N_L=58-63$),
$^{100-104}_{40}$Zr--$^{148-144}_{56}$Ba ($N_L=60-64$),  
$^{106-108}_{42}$Mo--$^{142-140}_{54}$Xe ($N_L=64-66$),  
$^{112}_{44}$Ru--$^{136}_{52}$Te ($N_L=68$),  and 
$^{116}_{46}$Pd--$^{132}_{50}$Sn ($N_L=70$).
There is also
one example of detected cold $^{10}$Be accompanied fission of $^{252}$Cf,
namely $^{96}_{38}$Sr--$^{146}_{56}$Ba ($N_L=58$).

On the right hand side of this figure there are plots for the new decay
modes which have a good chance to be
detected: 2$\alpha$-, 3$\alpha$-, and 4$\alpha$ accompanied fission.
The corresponding $Q$-values 
are not smaller compared to what has been already measured,
which looks very promising for the possibilty of detecting the 2$\alpha$-,
3$\alpha$-, and 4$\alpha$ accompanied fission decay modes. In fact by taking
into account the mass-values of the participants, one can see that the
$Q$-value for the 2$\alpha$ accompanied fission may be obtained by
translation with +0.091~MeV from the $Q$-value of the $^8$Be accompanied
fission. A similar translation with -7.275~MeV should be made from the
$^{12}$C accompanied fission in order to obtain the $Q$-values of the
3$\alpha$ accompanied fission, etc.
Less promising looks the combination of three cluster, $\alpha +
^6$He~+~$^{10}$Be accompanied cold fission of $^{252}$Cf. As mentioned
above, the $2\alpha$ accompanied fission was already observed.

\begin{figure}[htb]
\centerline{\includegraphics[width=12cm]{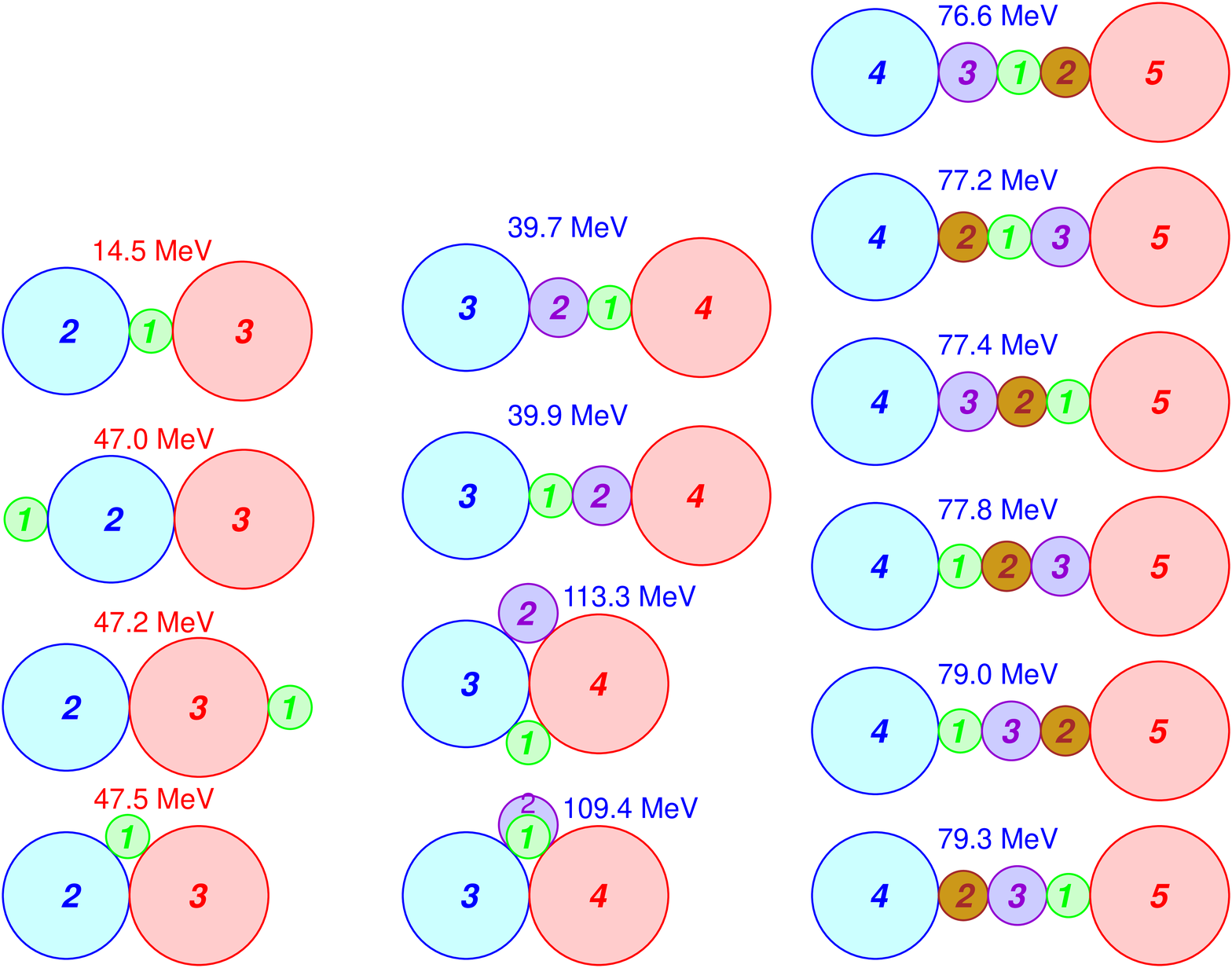}}
 \caption{Aligned and compact configurations for $\alpha$ accompanied cold
fission (left hand side) and
for $\alpha$~+~$^{10}$Be~accompanied cold fission (middle).
On the right hand side there are aligned configurations with three
clusters between the light and
heavy fragment for $\alpha$~+~$^6$He~+~$^{10}$Be~accompanied cold fission.
The parent is $^{252}$Cf, and the heavy fragment is the doubly magic
$^{132}$Sn. The corresponding energies are shown.
\label{fig11}}
 \end{figure}

Different kinds of aligned and compact configurations of fragments in touch 
are shown in figure~\ref{fig11}. On the left hand side
there are three aligned fragments on the same axis,
in the following order of the three partners: 213, 123, and 132 (or 231)
and one compact configuration (in which every partner is in touch with
all others). It is clear that the potential barrier for the ``polar
emission''
(123 or 132) is much higher than that of the emission from the neck (213),
which explains the  experimentally determined low yield of the polar
emission compared to the ``equatorial'' one. As it should be, the compact
configuration posses the maximum total interaction energy, hence it has the
lowest chance to be observed. 
The same is true for the quaternary fission when the two clusters are 
formed in the neck (middle part of the figure).
An important
conclusion can be drawn, by generalizing this result, namely: {\em the
multiple clusters $1, 2, 3, \ldots$ should be formed in a
configuration of the nuclear system in which there is a relatively long
neck between the light ($n-1$) and heavy ($n$) fragment.}
Such shapes with long necks in fission  have been considered~\cite{hil58c}
as early as 1958. For the ``true'' ternary fission, in two $^{84}$As
plus $^{84}$Ge, $E_t = 98$~MeV! Despite the larger $Q$-value
(266~MeV), the very large barrier height explains why this split has a
low chance to be observed.

On the right-hand side of figure~\ref{fig11}, we ignore the aligned configurations 
in which the heavy fragments are not lying at the two ends of the chain.
By arranging in six different manners the $\alpha$-particle, $^6$He, and
$^{10}$Be clusters between the two heavy fragments from the cold fission of 
$^{252}$Cf, the difference in energy is relatively small. Nevertheless, the
43125 configuration seems to give the lowest barrier height.

The energies of the optimum configuration of fragments in touch, 
for the 2$\alpha$-, 3$\alpha$-, and 4$\alpha$ accompanied cold fission
of $^{252}$Cf are not much 
higher than what has been already measured.
When the parent nucleus is heavier, the multicluster emission is stronger
as we observed by performing calculations for nuclei like 
$^{252,254}$Es, $^{255,256}$Fm, $^{258,260}$Md, $^{254,256}$No, 
$^{262}$Lr, $^{261,262}$Rf, etc.

While the minimum energy of the most favorable aligned configuration of
fragments in touch, when at least one cluster is not an alpha particle,
becomes higher and higher with increasing complexity of the partners, the
same quantity for multi alphas remains favorable.
In conclusion, we suggested since 1998 experimental searches for the
multicluster $2\alpha$ accompanied fission, 
for $^8$Be-,  $^{14}$C- and $^{20}$O accompanied fission. Also, the 
contribution of the single- and  multi-neutron accompanied cold fission 
mechanism to the prompt neutron emission has to be determined.

\section{Conclusions}
The method of finding the most general axially-symmetric shape at the 
saddle point without introducing {\em apriori} a parametrization (inherently
limited due to the finite number of deformation coordinates), by solving
an integro-differential equation was tested for binary, ternary, and
quaternary fission processes within a pure liquid drop model. 
The well known LDM saddle point shapes are well reproduced.
The method proved its practical capability in what concerns fission into two, 
three, or four identical fragments, for which
fission barriers given by shapes with rounded necks are, as expected, 
lower than those of aligned spherical fragments in touch.

Nevertheless, in the absence of any shell corection it is not possible
to reproduce the experimental data, or to give results for 
particle-accompanied fission. 
By adding (phenomenological) shell corrections  we succeded to obtain minima
at a finite value of mass asymmetry for the binary fission of 
$^{226-238}$Th and $^{230-238}$U nuclei.

Fission barriers for ternary and quaternary fission into identical fragments
are lower than for aligned spherical fragments in touch. Our expectations
concerning the possibilty to detect quaternary fission as
$2\alpha$-accompanied fission were experimentally confirmed.

\section*{Acknowledgements}
This work was partly supported by the Frankfurt Institute for Advanced
Studies, by a grant of the Deutsche Forschungsgemeinschaft, and by Ministry
of Education and Research, Bucharest.

\begin{appendix}

\section*{APPENDIX: Euler - Lagrange equation}

The functional to be minimized is the surface and Coulomb energy (given in
equation (\ref{eq:a})) with the constraints (eqs. \ref{eq:v}, \ref{eq:d}). We
denote with $F_1$, $F_2$, $F_3$, $F_4$, the corresponding integrands one
needs to write the Euler-Lagrange equation:
\begin{equation}
\sigma F_1 = \sigma \rho \left(1 + \rho'^2\right)^{1/2}
\end{equation}
\begin{equation}
\frac{R_0\rho _e\phi_s}{5} F_2 = \frac{R_0\rho _e\phi_s}{5} 
\left(\rho^2 - z\rho\rho'\right)
\end{equation}
\begin{equation}
 F_3 = \rho^2
\end{equation}
\begin{equation}
 F_4 = \rho^2 F
\end{equation}
The derivatives are easily obtained
\begin{equation}
\frac{\partial F_1}{\partial\rho}=\left(1+\rho'^2\right)^{1/2}
\end{equation}
\begin{equation}
\frac{d}{dz}\frac{\partial F_1}{\partial \rho'} =
\frac{d}{dz}\left[\frac{\rho\rho'}{(1+\rho'^2)^{1/2}}\right] =
\frac{\rho'^2+\rho\rho''}{(1+\rho'^2)^{1/2}} -
\frac{\rho\rho'^2\rho''}{(1+\rho'^2)^{3/2}} 
\end{equation}
\begin{equation}
\frac{\partial F_2}{\partial\rho}=2\rho -z\rho'
\end{equation}
\begin{equation}
\frac{d}{dz}\frac{\partial F_2}{\partial \rho'} = \frac{d}{dz}(-z\rho) =
-\rho - z\rho'
\end{equation}
\begin{equation}
\frac{\partial F_3}{\partial\rho}=2\rho 
\end{equation}
\begin{equation}
\frac{\partial F_4}{\partial\rho}=2\rho \left(F +
\frac{\rho}{2}\frac{\partial F}{\partial \rho}\right) 
\end{equation}
Consequently the Euler-Lagrange equation can be written as
\begin{equation}
\sigma\left(\frac{\partial F_1}{\partial\rho} - \frac{d}{dz}\frac{\partial
F_1}{\partial \rho'}\right) + \frac{R_0\rho _e\phi_s}{5}
\left(\frac{\partial F_2}{\partial\rho} -
\frac{d}{dz}\frac{\partial F_2}{\partial \rho'}\right) + \rho
\left(2\lambda_1'' + 2\lambda_2'' f\right) = 0  
\end{equation}
leading to 
\begin{equation}
\rho\rho''- \rho'^2- \left( \lambda_1 + \lambda_2|z| +
6XV_s \right)\rho (1+\rho'^2)^{3/2} - 1 = 0
\end{equation}
if we choose $F=|z|$ (hence $f=|z|$) and express $3R_0\rho_e/(5\sigma)$ as
$6X$ because the Coulomb
and surface energy of a spherical nucleus within LDM are given by
$E_C^0=(3Z^2e^2)/(5R_0)$ and $E_s^0=4\pi R_0^2\sigma$, respectively.

Alternatively one can obtain from this equation the equivalent
relationship
\begin{equation}
2\sigma K + 3\rho_e \phi_s/5 + {\lambda}'_1 + {\lambda}'_2 f = 0
\; \; ; \; \; f(z,\rho) = F(z,\rho) + \frac{\rho}{2}\frac{\partial
F(z,\rho)}{\partial \rho}
\end{equation}
in which the definition of the mean curvature, $K$, from section \ref{equ} was
used.

\end{appendix}

\newpage

\bigskip 

\newpage

\bigskip \bigskip

\newpage

\bigskip \bigskip

\newpage

\bigskip \bigskip 

\newpage

\bigskip \bigskip \bigskip \bigskip

\newpage

\bigskip \bigskip \bigskip \bigskip

\newpage

\bigskip \bigskip \bigskip \bigskip

\newpage

\bigskip \bigskip \bigskip \bigskip

\newpage

\vspace*{4cm}

\newpage

\renewcommand{\baselinestretch} {0.8}

\begin{table}
\caption{\label{table3} Conditional saddle point elongation parameter 
$d_{R-SP}$ and energy $E_{SP}$ for a given asymmetry $d_L-d_R$ within a 
pure liquid drop model for $^{238}$U parent nucleus.}
\begin{center}
\begin{tabular}{|ccc|} \hline
 $d_{R-SP}$ & $d_L-d_R$ & $E_{SP}$ (MeV) \\   
\hline 
    1.405&0.000&5.822\\
    1.400&0.005&5.840\\
    1.400 & 0.010 &5.889\\
    1.395 & 0.015 &5.972\\
    1.410 & 0.020 &6.029\\
    1.395 & 0.025 &6.237\\
    1.390 & 0.030 &6.420\\
    1.390 & 0.035 &   6.637 \\
    1.390 & 0.040 &   6.888 \\
    1.390 & 0.045 &   7.172 \\
    1.390 & 0.050 &   7.492 \\
%    1.390 & 0.055 &   7.845 \\
    1.390 & 0.060 &   8.232 \\
%    1.390 & 0.065 &   8.657 \\
    1.390 & 0.070 &   9.115 \\
%    1.395 & 0.075 &   9.612 \\
    1.395 & 0.080 &  10.144 \\
%    1.400 & 0.085 &  10.712 \\
%    1.400 & 0.090 &  11.317 \\
%    1.430 & 0.095 &  11.815 \\
\hline 
\end{tabular}
\end{center}
\end{table}

\newpage

\begin{table}
\caption{\label{table4}Conditional saddle point elongation parameter 
$d_{R}$ and energy $E_{SP}$ 
for a given asymmetry $d_L-d_R$ within liquid drop model plus
phenomenological shell correction for $^{238}$U
parent nucleus. The shell correction for the spherical parent $\delta E^0 =
-5.385$~MeV. For $d_L-d_R < 0.045$ the $d_R$ at which $E_{SP}$ is maximum
differs from $d_R$ at which its LDM part is maximum.}
\begin{center}
\begin{tabular}{|ccccc|}  \hline
 $d_R$ & $d_L-d_R$ & $E_{SP-LDM}$ & $\delta E_{SP} - \delta E^0$ &
$E_{SP}$ \\   
   &  &MeV  &MeV  &MeV  \\
\hline 
    1.385 & 0.000 & 5.728 & 0.462 & 6.189 \\
    1.390 & 0.000 & 5.775 & 0.413 & 6.188 \\
\hline
    1.385 & 0.005 & 5.769 & 0.410 & 6.179 \\
    1.410 & 0.005 & 5.823 & 0.286 & 6.109 \\
\hline
    1.395 & 0.010 & 5.804 & 0.345 & 6.150 \\
    1.405 & 0.010 & 5.859 & 0.250 & 6.109 \\
\hline
    1.385 & 0.015 & 5.936 & 0.146 & 6.082 \\
    1.390 & 0.015 & 5.841 & 0.282 & 6.123 \\
\hline
    1.385 & 0.020 & 6.062 &-0.064 & 5.998 \\
    1.390 & 0.020 & 5.953 & 0.129 & 6.082 \\
\hline
    1.385 & 0.025 & 6.219 &-0.327 & 5.892 \\
    1.400 & 0.025 & 6.108 &-0.161 & 5.948 \\
\hline
    1.385 & 0.030 & 6.408 &-0.642 & 5.766 \\
    1.395 & 0.030 & 6.233 &-0.375 & 5.858 \\
\hline
    1.385 & 0.035 & 6.628 &-1.010 & 5.618 \\
    1.395 & 0.035 & 6.421 &-0.665 & 5.755 \\
\hline
    1.385 & 0.040 & 6.881 &-1.431 & 5.450 \\
    1.395 & 0.040 & 6.634 &-0.996 & 5.639 \\
\hline
    1.415 & 0.045 & 7.044 &-1.201 & 5.844 \\
    1.405 & 0.050 & 7.445 &-0.289 & 7.155 \\
    1.400 & 0.055 & 7.825 & 0.541 & 8.366 \\
    1.395 & 0.060 & 8.229 & 1.296 & 9.525 \\
    1.390 & 0.065 & 8.657 & 2.403 & 11.059 \\
\hline 
\end{tabular}
\end{center}
\end{table}

\end{document}